# Hosting Capacity Approach Implications

Narayan Bhusal*, Andrija Sadikovic**, and Mohammed Ben-Idris*
* Department of Electrical & Biomedical Engineering
University of Nevada, Reno, Reno, NV 89557
** Quanta Technology, LLC, one Concord Centre, California

## SUMMARY

This paper revisits the generation hosting capacity (HC) calculation approach to account for grid operational flexibility—the ability to reconfigure the system safely. In essence, the generation hosting capacity is determined against the set of limiting factors—voltage, thermal (conductor loading), reverse flow (at the feeder head, station transformer, or substation), and change in the voltage (due to sudden change in generation output). Not that long ago, California Investor-Owned Utilities (IOUs) added a new criterion that does not allow reverse flow at the supervisory control and data acquisition (SCADA) points that can change the system configuration, aiming to prevent the potential transfer of reverse flow to an adjacent feeder. This new criterion intended to capture operational constraints as part of hosting capacity—known as hosting capacity with operational flexibility (OpFlex). This paper explores the shortfalls of such an approach and proposes performing actual transfer analysis when determining hosting capacity rather than implementing the OpFlex approach. Furthermore, we discuss the need for transition to determining hosting capacity profile (all intervals) rather than a flat line (one, worst performing interval) hosting capacity. A hosting capacity profile would inform the developers of interval-by-interval limits and opportunities, creating new opportunities to reach higher penetration of DERs at a lower cost. With technological and computational advancements, such an approach is neither out of implementation reach nor that computationally expensive. In return, far more DER can be interconnected once programmed not to violate certain generation profiles as part of the interconnection requirement, and utilities would be better informed of their actual operational flexibility, benefiting society overall.

bhusalnarayan62@nevada.unr.ed

# 1 INTRODUCTION

Generation hosting capacity (HC) is the amount of uniform generation output integrated at a specific part of a distribution system without violating operational and control limits. The thermal, voltage (upper and lower voltage limit and voltage deviation), and protection criteria have been implemented in the distribution system's hosting capacity analysis [1]. The electric utility industry has emphasized the need to account for operational flexibility (OpFlex) as part of hosting capacity analysis—and California Investor-Owned Utilities (IOUs) have initially attempted to approximate OpFlex [7]. This simplified OpFlex based hosting capacity approach restricts zero power flow limit through any supervisory control and data acquisition (SCADA) controlled load transfer devices in addition to thermal compliance, voltage compliance, protection device compliance, and no reverse power flow at the feeder/transformer/substation [6]. The Integrated Capacity Analysis (ICA) working groups in California adopted this concept and methodology as part of the public capacity maps published by each IOU.

The criteria of zero reverse power flow through the SCADA-controlled load transfer devices is intended to limit the transfer of system reverse flow to adjacent feeders. OpFlex criteria restricting reverse power flow at the feeder's head imposes potentially lower limits compared to other hosting capacity limits. In addition, zero reverse power flow through the SCADA-controlled load transfer devices criterion imposes even more stringent limitations to hosting capacity [4].

Taking the worst-case scenario approach in determining hosting capacity can significantly reduce the hosting capacity results. Deboever et al. propose using a set of scenarios that capture worst-case conditions and associated probability outcomes to determine hosting capacity [2]. They point out that using peak and min demand profiles does not always capture the worst-case conditions. The selection and use of load profiles are very important aspects of the hosting capacity process. This statement is particularly true when low probability conditions are jointly applied, potentially using an inaccurate approach.

Hosting capacity has an important impact on our society with enormous long-term implications for clean energy goals and electric grid reliability and resilience. Therefore, a hosting capacity approach must be carefully designed and reinforced by regulatory actions to maintain a sustainable alignment of societal goals and Investor-Owned Utilities (IOUs) business needs. This paper revisits existing industry-practiced hosting capacity approaches, specifically, (a) industry-practiced OpFlex-based hosting capacity calculation approach and (b) worst-case scenario-based hosting capacity approaches. On the contrary, the profile-based interval hosting capacity concept is introduced with justifications provided through case studies. Similarly, advocating for analysing the actual load transfer through network reconfiguration rather than simply considering the zero reverse power flow limit through SCADA-controlled load transfer devices captures true operational flexibility as part of hosting capacity. In this work, we have also performed case studies with different DER penetration levels, specifically photovoltaic (PV) and electric vehicles (EVs), to better evaluate the potential implications of a hosting capacity approach in the future.

The rest of the paper is arranged as follows: 1) Section 2 describes the trade-offs of the existing hosting capacity calculation approaches. 2) Section 3 provides the system modelling. 3) Section 4 provides the implication of various hosting capacity approaches. 4) Finally, Section 5 contains concluding remarks.

## 2    UNDERSTANDING TRADEOFFS

California IOUs practice a deterministic HC approach with quasi-operational flexibility limit (a.k.a. OpFlex) criteria—no reverse flow at any time is allowed if a device is subject to load transfer to an adjacent distribution circuit. Although from a calculation and practical implementation perspective this approach raises several concerns:

1. **Uniform Value vs. Profile**: A worst-case interval approach imposes a uniform limit across all intervals, eliminating DER opportunities at other, non-limiting intervals. By publishing HC profiles, DERs can be (self) controlled or supplemented with energy storage not to exceed specified net outputs (HC profiles) at specified intervals.
2. **Extreme vs. Expected Outcome**: OpFlex is based on low probability outcomes. The approach assumes low probability (peak and minimum) demand profiles during the load transfer (abnormal system conditions). The minimum and peak demand represents the $10^{th}$ and $90^{th}$ percentile of historic demand, respectively—low probability outcomes. The probability of such extreme conditions occurring during periods of an abnormal condition is very low. Designing and maintaining a system for very low probability outcomes can be costly, outdated, and overly simplified.
3. **Actual vs. Approximated Operational Flexibility**: Operational flexibility is not well captured with OpFlex, which attempts to approximate operational flexibility without performing actual transfer studies. The current approach is likely to over- or under-estimate HC, potentially posing a risk to distribution system performance. Furthermore, while OpFlex implementation is simple in the analysis, it fails to capture DER implications during load transfers. Specifically, it is common to experience a fault or have scheduled maintenance in part of the system, requiring system reconfiguration. Otherwise, the system remains in its normal configuration. As implemented, OpFlex does not consider the loss of de-energized sections. One can expand this further by considering the partial loss of the feeder—such as blown lateral fuse—without initiating load transfer. Hence, one should understand trade-offs when seeking simplicity in the implementation process.
4. **General vs. Location Specific Reverse Flow Limit**: Global implementation of no reverse flow limit may prevent DER penetration if no adverse impact occurs during load transfer activities. There should be only one zero reverse flow limit—feeder head, station transformer, or substation—that imposes a limitation on DER penetration for all normal and abnormal system configurations. As a global criterion, it results in many sections with zero generation HC results, regardless of whether adjacent feeder(s) can absorb such reverse flow. As part of HC studies, transfer analysis is more time demanding, but the upside is potentially too large not to do it.
5. **Model with vs. without Inverter Setting**: Smart inverter functions are typically not modelled in the HC analysis. Although this is more of an industry issue, it is worth mentioning that smart inverter functions must be modelled to determine the proper HC results.

Section 4 cannot detail all of these concerns because of article page limits. Attorneys for Interstate Renewable Energy Council Inc. have also pointed out some of these concerns [6]. However, it is important to consider all these points when crafting an HC approach.

## 3    SYSTEM MODELLING

This work used Synergi to determine incremental HC at every section on a distribution feeder with existing loads. Analysis was repeated for different photovoltaic (PV) and electric vehicle (EV) penetration levels, where a 0% PV and 0% EV scenario represents the base case. For each penetration scenario, the deterministic approach (incremental HC approach, the most common industry-based HC calculation approach) provided HC results for normal and abnormal (load transfer from/to adjacent feeders) system conditions. In this paper, we have not included the rigorous description of the incremental HC approach, interested readers are refer to [8] and reference therein.

We analysed several feeder pairs to evaluate the current approach's implications, including load transfers by opening and closing appropriate circuit switches. The findings point to the same conclusions, but in this paper, we show study results for a feeder pair with the following characteristics:

- F1 – 1376 sections, 11.3/5.6 MW peak/min demand MW peak load, 62. 2 conductor miles long(miles), and 1306 connected customers.
- F2 – 825 sections, 9.4/1.2 MW peak/min demand MW peak load, 48.7 conductor miles long(miles), and 1382 connected customers.

To assess the potential implications of the existing approach in future years, we modelled several scenarios with different photovoltaic (PV) and light-duty electric vehicle (EV) penetration levels, as shown in Table 1.

**Table 1. PV and EV Penetration Levels.**

|        | PV 0% | PV 20% | PV 40% |
|--------|-------|--------|--------|
| EV 0%  | x     | x      | x      |
| EV 20% | x     | x      | x      |
| EV 40% | x     | x      | x      |

## 3.1 Light-duty Electric Vehicles

Light-duty EV charging was modelled using outputs of the Electric Vehicle Infrastructure Projection Tool (EVI-Pro) Lite developed by the National Renewable Energy Laboratory (NREL) [3]. This tool estimates EV charging profiles based on geographical location (state and city). Several factors have been considered while calculating the EVs profiles, including 1) miles travelled per vehicle daily, 2) average ambient temperature, 3) the ratio of all-electric EVs and hybrid vehicles, 4) vehicle class (e.g., sedan and sport utility vehicle), 5) preferred charging location (home or workplace), 6) access to charging at home, 7) the ratio of home vs. workplace charging, and 8) charging strategies (e.g., immediate and delayed). Figure 1 shows the power demand profile for weekdays and weekends to charge 1,000 electric vehicles with an average daily travel distance of 25 miles per vehicle at a daily average temperature of 68-degree Fahrenheit in Raleigh, North Carolina. Other parameters used to generate Figure 1 are as follows: 1) 50% of all vehicles are all-electric (another 30% are hybrid-electric), 2) 80% of the vehicles are sedan, 3) 20% of the chargers in the workplace are level 1 (1.4 kW) while remaining are level 2 (6.2 kW), 4) all vehicles have access to home charging with 50% level 1 and 50% level 2 chargers, 5) home is the preferred charging location, and 6) the home and workplace charging strategies are immediate (as fast as possible). For details on the electric vehicle demand projection, we refer to [3].

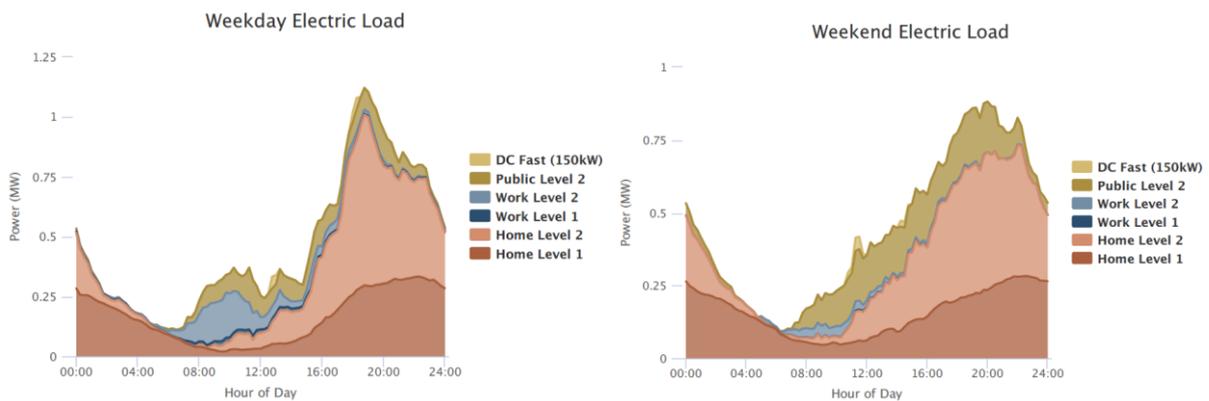

**Figure 1. Weekday and Weekend Electric Vehicles Charging Profile in Raleigh, North Carolina.**

Once the EVI-Pro dataset described above was generated, charging behaviour was adjusted for monthly/hourly temperature variations and seasonal (summer and winter) average daily driving range to model changing driving behaviours based on the seasons (average driving). The average daily driving range assumed in this study is as shown in Table 2. The number of modelled EV chargers is lower than the number of light-duty EVs adopted on a distribution circuit, representing actual charging activity that in aggregate matches the overall demand of EVs seen by the distribution system. Specifically, for 1,000 EV vehicles that experience an average daily driving distance of 45 miles, the distribution of EV charger types is as follows: 331 home level 1, 132 home level 2, 30 work level 1, 79 work level 2, 82 public level 2, and 2 DC fast chargers.

**Table 2. Daily Average Driving Distances.**

|  | Jan | Feb | Mar | Apr | May | Jun | July | Aug | Sep | Oct | Nov | Dec |
|---|---|---|---|---|---|---|---|---|---|---|---|---|
| Daily Miles | 24 | 24 | 26 | 26 | 26 | 28 | 28 | 28 | 26 | 26 | 26 | 24 |

### 3.2    Photovoltaic System

PV profiles were determined using the PVWatts Calculator [5]. Commercial and residential site maximum daily output for January was approximately 30 and 2.5 kW, respectively. We have not provided the PVWatts Calculation tool's details in this paper, but those interested, interested reads are referred to [5].

### 3.3    Switching

Six mainline switching blocks were identified for this feeder pair. Therefore, four switching configurations were evaluated, plus the base case (normal configuration) scenario.

## 4    APPROACH IMPLICATIONS

### 4.1    Operational Flexibility

Operational HC limits are not well captured with OpFlex criteria implemented in California. California IOUs' intention to capture operational aspects as part of HC analysis is very important and needed. However, the difference in HC results between the two approaches—actual load transfer and OpFlex—is significant. Figure 2 and Figure 3 show aggregate HC across sections at specific distances for 3 and 1&2 phase sections, respectively.

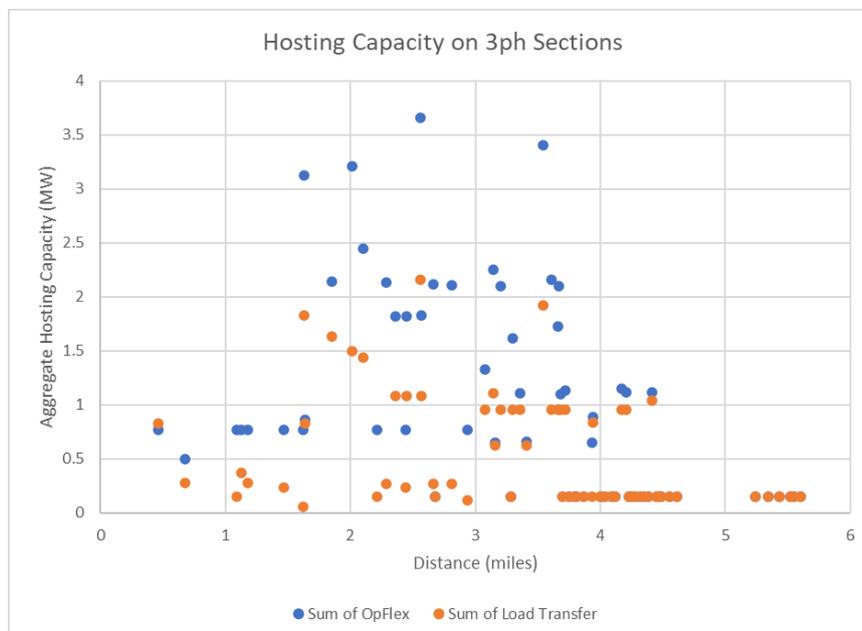

**Figure 2. Aggregate Generation HC Results for 3 Phase Sections by Distance from the Substation.**

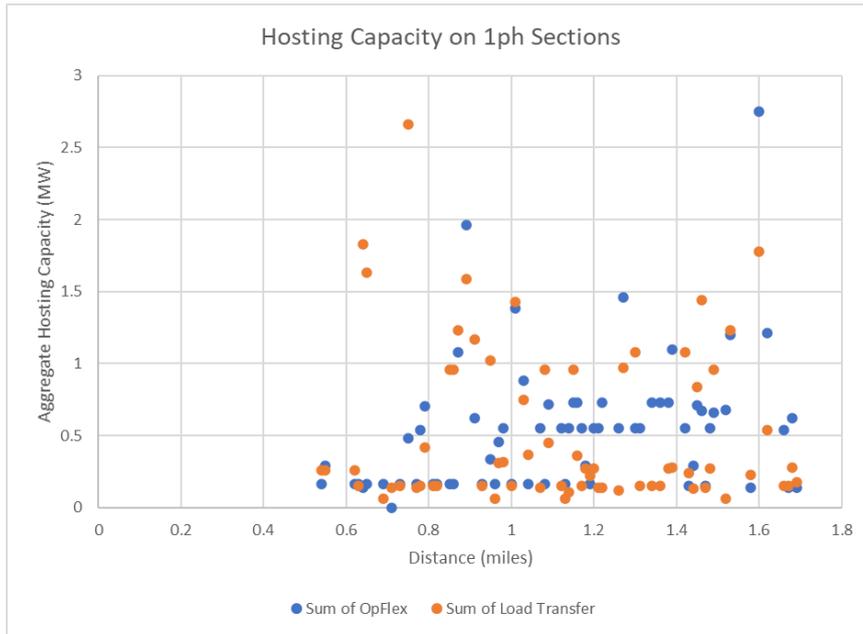

**Figure 3. Aggregate Generation HC Results for 1-2 Phase Sections by Distance from the Substation.**

When evaluating the implications of an HC approach, it is important to observe which limits are most frequent across different system configurations. As shown in Figure 4, the HC limits distribution is different between the existing (OpFlex) approach and configuration scenarios (base case and configurations 1 through 4). With the OpFlex limit removed, the resulting hosting capacity limits still capture the worst-case outcome across the base case and four abnormal configurations. To clarify, we are not discussing probabilities associated with these outcomes as it is outside of this paper's scope.

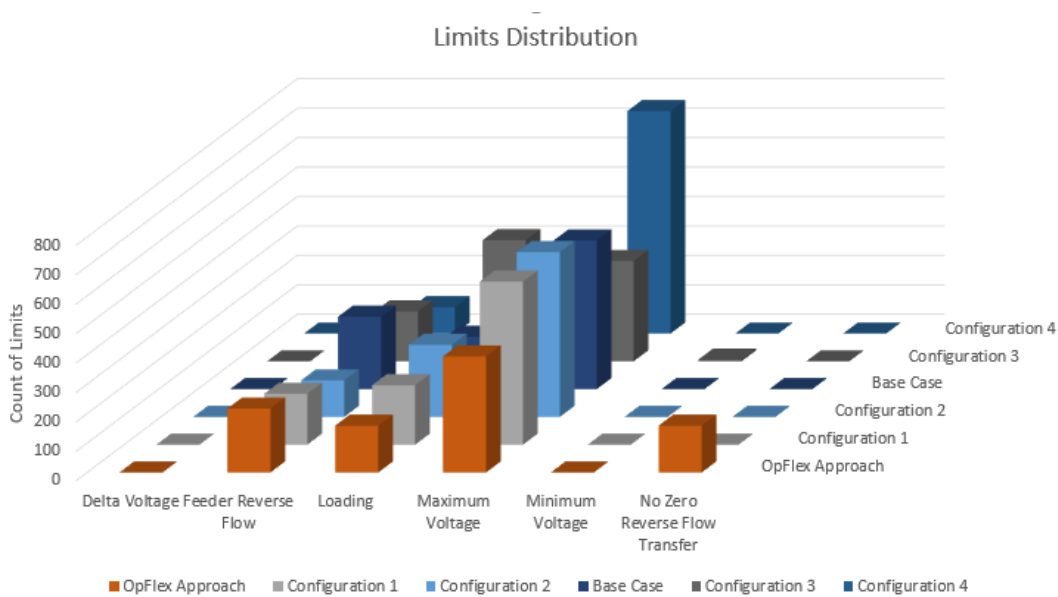

**Figure 4. Generation HC Constraining Limits Distribution Comparaison.**

As operational flexibility has been identified as a potential limiting capacity analysis component, the current implementations often become limited by minimum loading conditions. The minimum load period where estimated DER generation would be highest causes an increased risk of reverse power flow. These assumptions inevitably impact grid performance for policy goals, such as EV and PV adoption. Figure 5. Difference in Generation HC Results per Section, OpFlex vs. Load Transfer Approach. shows the difference in HC results between the existing (OpFlex) and HC analysis based on load transfer configurations. The difference is both positive and negative in value and varies across the sections and penetration scenarios.

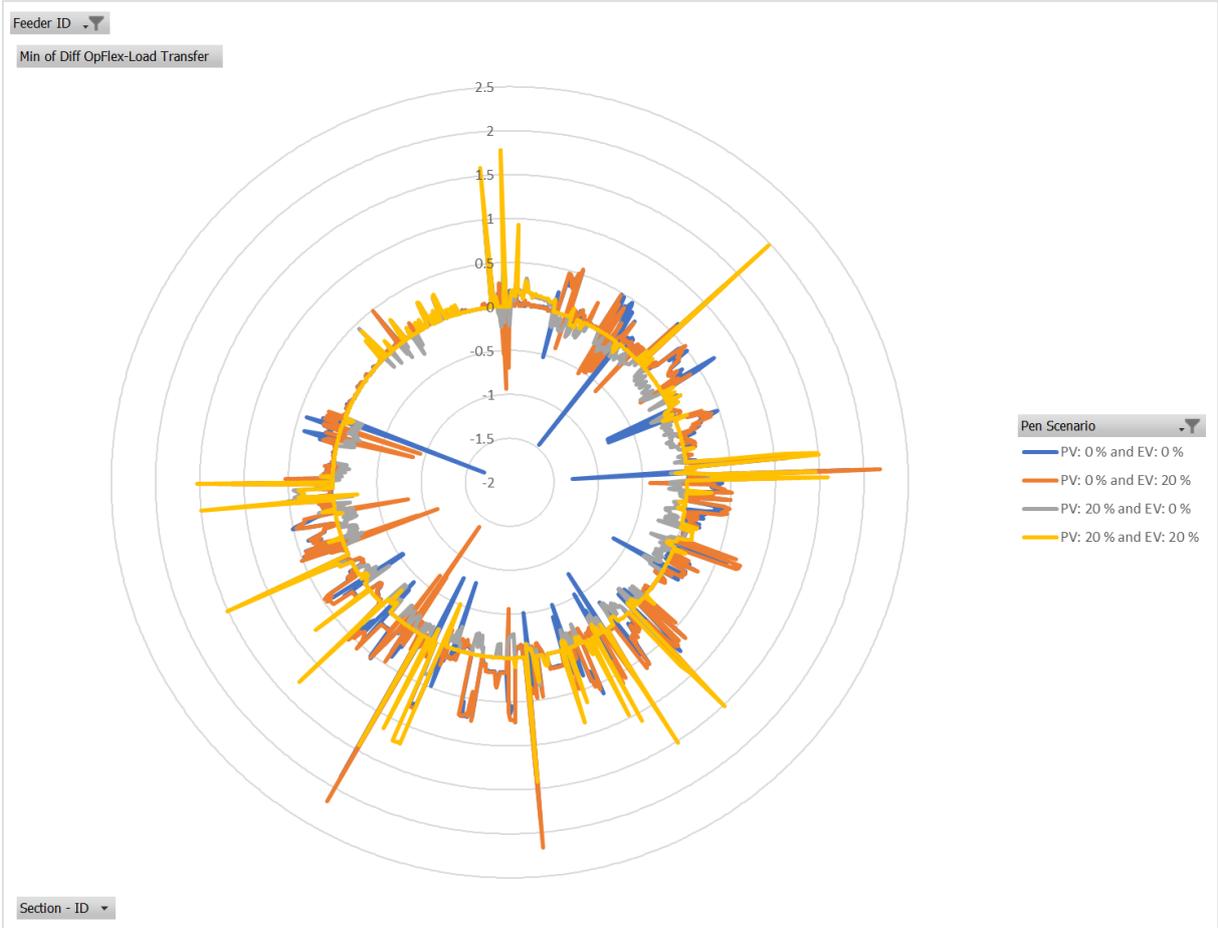

**Figure 5. Difference in Generation HC Results per Section, OpFlex vs. Load Transfer Approach.**

OpFlex over-estimates HC even at higher EV penetration levels. Figure 6 shows the difference between the OpFlex and the load transfer approaches for aggregated HC values across both study feeders. There is a nearly consistent positive difference across all phasing configurations for almost all penetration level combinations. These results may vary based on locations where PV and EV adoption occur. Regardless, the difference is too significant to ignore the shortfall of the OpFlex approach.

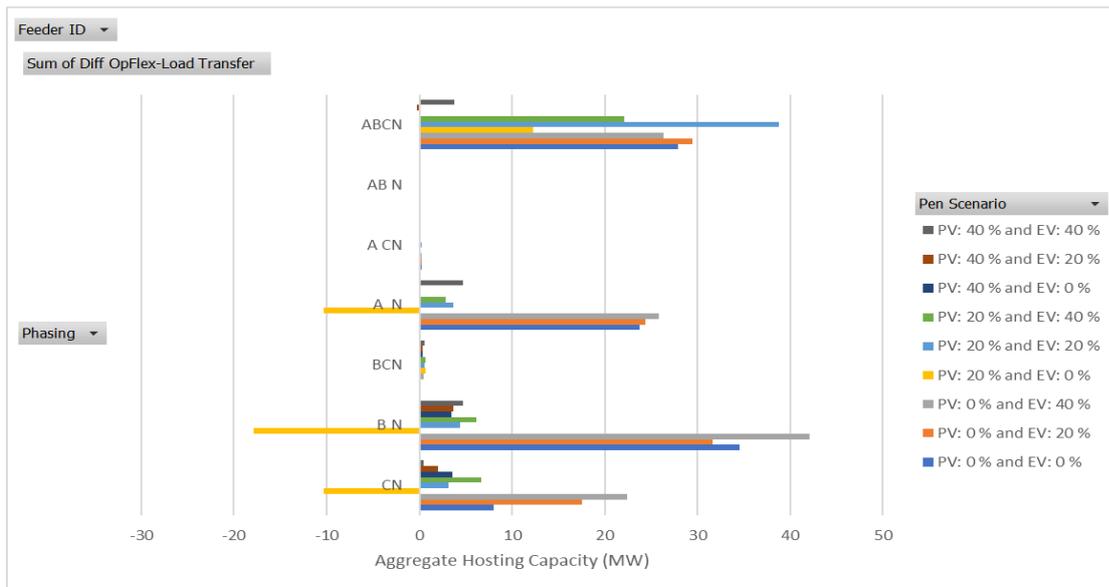

**Figure 6. Difference in Aggregate Generation HC between OpFlex and Load Transfer Approaches.**

This difference leads to two questions: 1) How well are we engineering our business and technical processes for a safe and reliable electric grid under high DER penetration? 2) How can we reach higher penetration levels if we impose strict limits providing less and less HC? Perhaps the answers lie in shifting the focus on using HC profiles instead of one value and/or the use of energy storage.

As we look further at load HC, there is little operational flexibility, even today. As shown in Figure 7, base case results are more dispersed with the least number of zero load HC sections. As soon as transfers occur, the number of zero load HC sections increase—for some configurations, 2/3 of sections end up with zero load HC. The results are consistent across all other penetration scenarios. These results imply that operational flexibility for hosting load at peak demand could decrease as we transfer load for operations. While this is not surprising, as many systems are designed to support normal—not necessarily abnormal—operating conditions at peak demand, operating flexibility is not planned and designed for peak demand. This situation raises the question: Should we apply multiple configurations to hosing capacity studies? The answer is yes, however, operational flexibility in HC studies requires the distribution system to be planned and designed for operational flexibility.

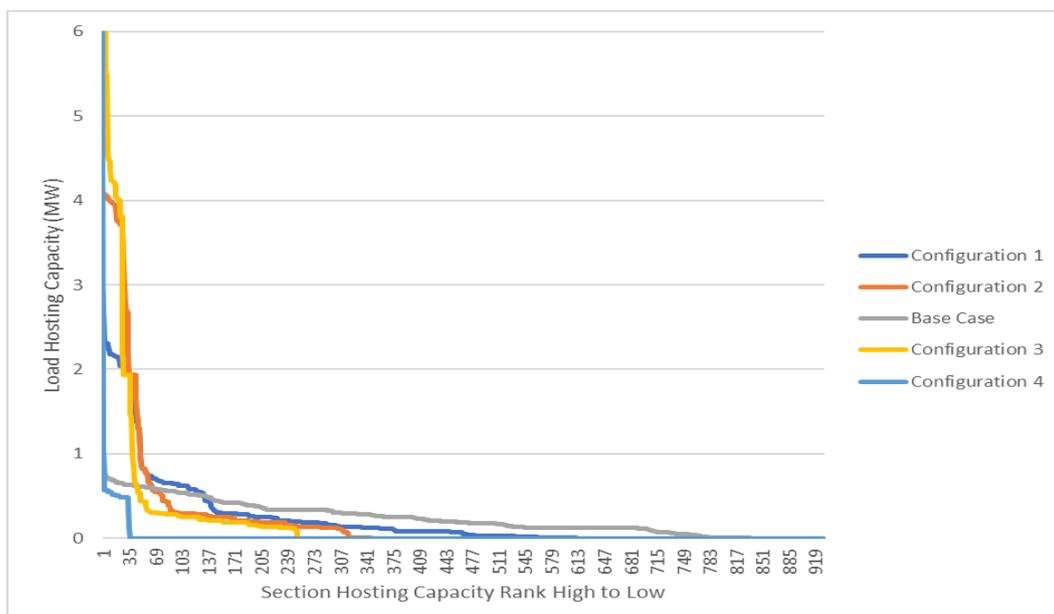

**Figure 7. Load HC Results for Different System Configurations: EV 0% PV 0% Penetration.**

## 4.2 Hosting Capacity Profile

HC profiles are more informative and beneficial to utilities and DER developers than the single-limit uniform HC results generated today. Limiting DER interconnection based on the worst outcome in one/any interval significantly lowers the amount of DER that could be interconnected given today's technological advancements. The amount of DER output that is (potentially) lost due to current approaches cannot be ignored, given the high clean energy goals across the country, including those in California.

To demonstrate the consequence of limiting DER output based on the worst-case interval, we show several plots of HC results of a sample feeder for individual intervals (between 6 AM and 6 PM for each month in a calendar year) in Figure 8. These plots show how much DER outputs would be lost—everything above the lowest point on the x axis—if DER output is limited to the lowest value (worst-case interval). Darker colours indicate more data points at that value.

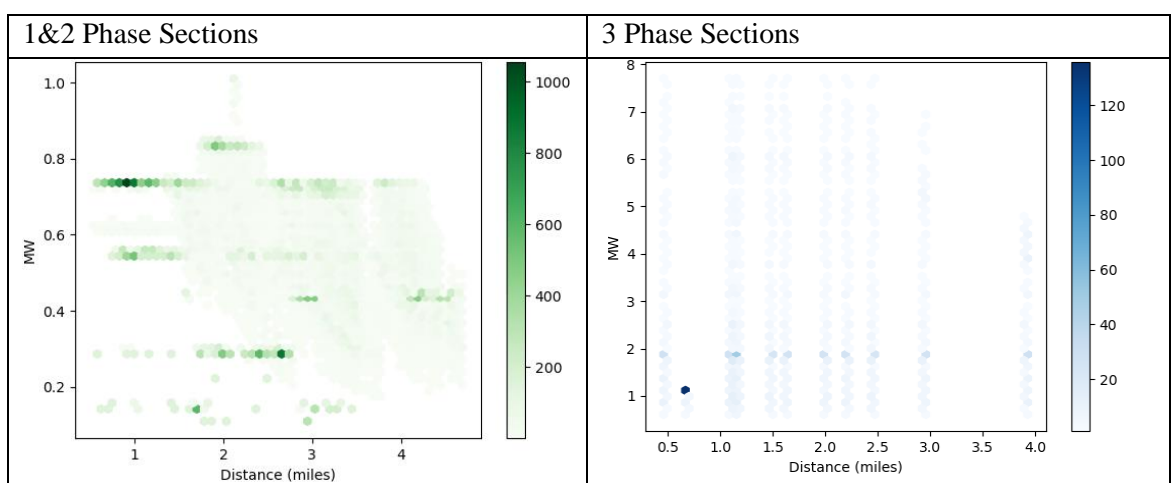

**Figure 8. Peak Day/Minimum Day Interval Based Generation HC at Different Intervals.**

## 4.3 Aligning Objectives

The more load transfer points on the system, the better chance of improving system reliability, the more constrained generation HC using the OpFlex approach. These are the opposing forces from the DER penetration perspective. Of course, the operational flexibility aspect should be integral to HC and to planning and operations activities—distribution system planning activities often focus on allowing planning margins to support abnormal system configuration, except for downtown networks. Using a consistent approach across planning and operational activities, including interconnection, would be extremely beneficial and non-discriminatory.

When the probability of such outcomes is considered, the OpFlex rationale becomes less reasonable. The probability of a load transfer (OpFlex event) happening during a low probability demand profile at or around a limiting interval is low. Such an approach seems too risk-averse, disallowing a large volume of DERs predominantly because of a probability of a reverse flow condition on a load transfer point on the system. Basing the analysis on a combination of low probability events—extreme demand profile and switching events occurring simultaneously—should be reconsidered. Perhaps, using weekend/weekday (historic average) profiles as an expected outcome is more appropriate for transfer studies—weighted for probability of switching configuration occurrence (expected outcome).

## 5 CONCLUSIONS

Existing implementations of operational flexibility calculations in California are prohibitively conservative, limiting published hosting capacity and further DER penetration. Whichever path regulators and utilities choose moving forward, it is important to understand system design and the HC approach's impact on policy goals. We recommend some level of load transfer analysis as part of the HC process rather than an approximation to calculate and publish the HC profile (not just a single value) and to explore opportunities for DERs to interconnect if DER output does not violate the HC limit profile rather than one interval uniformly applied limit. In other words, calculate and allow DERs interconnection of profile shapes that are self-enforceable at the DER level. Also, we recommend that utilities standardize the planning and DER interconnection process, including HC calculation. Operational flexibility should be planned and considered systematically across impact studies, including HC. Lastly, if operational limits are considered, HC should consider the probability of such outcomes rather than worst-case outcomes regardless of probability. A key long-term solution in overcoming HC constraints is using utility-scale and grid-edge energy storage.

## ACKNOWLEDGMENTS

The United States National Science Foundation supported this work under grant number 1847578. The authors thank Brad Jensen and Stephen Steffel from Quanta Technology for providing valuable comments. The authors are responsible for all the statements and the conclusions made in this work.